\documentclass[a4paper]{article}

\usepackage{atmohead2013}
\usepackage[english]{babel}
\usepackage{enumitem}
\usepackage{dblfloatfix}

\title{Muon efficiency of the H.E.S.S. telescope}

\shorttitle{H.E.S.S. muon efficiency}

\authors{
R. Chalme-Calvet$^{1}$,
M. De Naurois$^{2}$,
J.-P. Tavernet$^{1}$,
for the H.E.S.S. Collaboration.
}

\afiliations{
$^1$ LPNHE, Universit\'e Pierre et Marie Curie Paris 6, Universit\'e Denis Diderot Paris 7, CNRS/IN2P3 \\
$^2$ LLR, Ecole Polytechnique, CNRS/IN2P3
}

\email{raphael.chalme-calvet@lpnhe.in2p3.fr}

\abstract{The H.E.S.S. cameras require a precise and regular calibration over time, to reconstruct the gamma-ray characteristics. The different sub-systems used to determine the gain and the uniformity of the PMTs and their evolution with time are presented. Then, we focus on the absolute energy scale calibration, by using a full reconstruction of isolated muons recorded during normal observation. The method and the evolution of the absolute overall light collection efficiency are shown.}


\begin{document}
\maketitle


\section{Introduction}

The H.E.S.S. experiment, located in Namibia, is a system of Imaging Atmospheric Cherenkov Telescopes dedicated to the detection of very high energy gamma-rays. It consists of five telescopes including four medium-sized telescopes (radius $\sim$ 6.5 m) and a large telescope (radius $\sim$ 15.5 m). Thereafter, we will focus on the medium-sized telescopes. A detailed description can be found in \cite{bib:calib}.

To analyse the observed gamma-ray characteristics, an accurate calibration of the electronic response of the 960 photomultiplier tubes (PMTs) of each H.E.S.S. camera as well as the evaluation of the instrument optical efficiency need to be performed.

The different steps are presented and the method to determine the optical efficiency is described in detail.

\section{H.E.S.S. cameras calibration}

The calibration of the H.E.S.S. cameras provides the conversion parameters from ADC counts to photoelectrons for each PMT.

In order to have a large dynamical range in intensity, the H.E.S.S. camera electronic is divided into two different channels: high gain (HG) and low gain (LG) channels. The amplitude in a PMT  in photoelectrons (p.e.) for both channels ($A^{HG}$ and $A^{LG}$) as a function of the amplitude in ADC counts ($ADC^{HG}$ and $ADC^{LG}$) is given by the following equations:
\begin{equation}
\left\{   
\begin{array}{l l}
A^{HG} = \frac{ADC^{HG} - P^{HG}}{\gamma_e^{ADC}} \times FF \\\\
A^{LG} = \frac{ADC^{LG} - P^{LG}}{\gamma_e^{ADC}} \times \frac{HG}{LG} \times FF
\end{array}
\right.
\end{equation}
where $\gamma_e^{ADC}$ is the gain of the high gain channel (in ADC/p.e.), $P^{HG}$ and $P^{LG}$ are the ADC positions of the base-line for both channels (pedestal position), $\frac{HG}{LG}$ is the amplification ratio of high gain to low gain and $FF$ is the flat-field coefficient of the studied pixel.

The determination of each of these parameters will be described in this part.

   \begin{figure}[ht]
  \centering
  \includegraphics[width=0.42\textwidth]{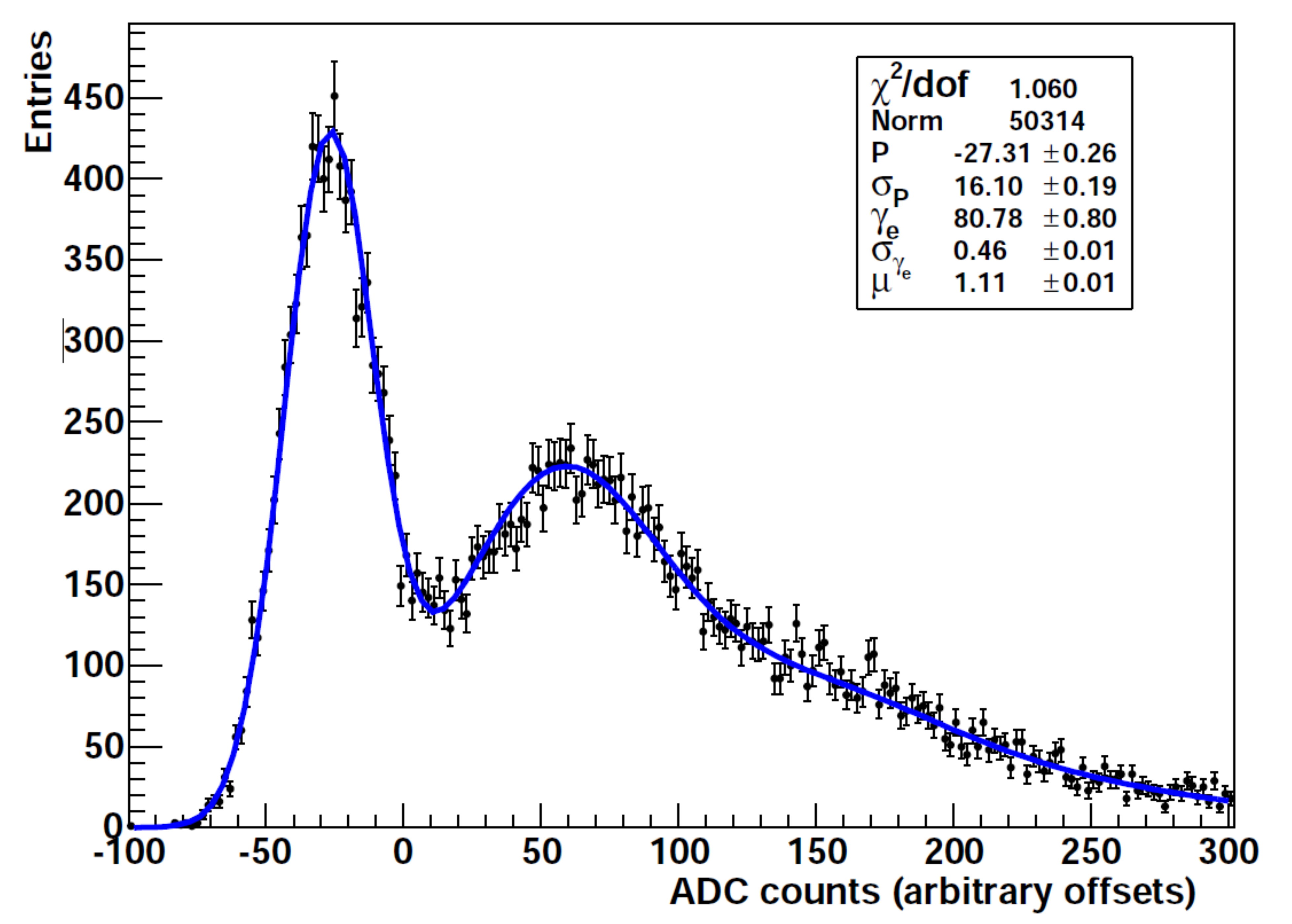}
  \caption{Single photoelectrons signal distribution (black points). The fitted function is represented in blue.}
  \label{fig:gain}
 \end{figure}

\subsection{Gains}

In order to determine the high gain, the camera in nominal configuration is illuminated by a pulsed blue light source (near one p.e. in average) in the dark during specific runs. The high gain $\gamma_e^{ADC}$ is then extracted from a fit of the signal distribution of each pixel in ADC counts units. An example of this distribution is shown in the figure \ref{fig:gain}. 

This signal distribution can be described by the following sum of Gaussian distributions:
\begin{eqnarray}
\mathcal{G}(x) = N \times \left(
\frac{e^{-\mu}}{\sqrt{2\pi} \sigma_{P}}
\exp\left[-\frac{1}{2} \left(\frac{x-P}{\sigma_P}\right)^2 \right]\right. \quad\quad\quad \nonumber \\
+\left. \kappa \sum_{n=1}^{m \gg 1} \! \frac{\mu^n e^{-\mu}}{\sqrt{2\pi} \sigma_{\gamma_{e}}n!}
\frac{}{} \!
\exp\left[- \left(\frac{x-(P+n\gamma_e^{ADC})}{\sqrt{2n}\sigma_{\gamma_{e}}}\right)^2 \right] 
\right)
\end{eqnarray}

The value of the low gain channel is then determined by looking at a specific range of intensity (between 30 and 150 p.e.) where the two channels are linear. In this range, the ratio of the two channel signals in ADC counts is directly equal to the amplification ratio $\frac{HG}{LG}$.

 \begin{figure}[th]
  \centering
  \includegraphics[width=0.42\textwidth]{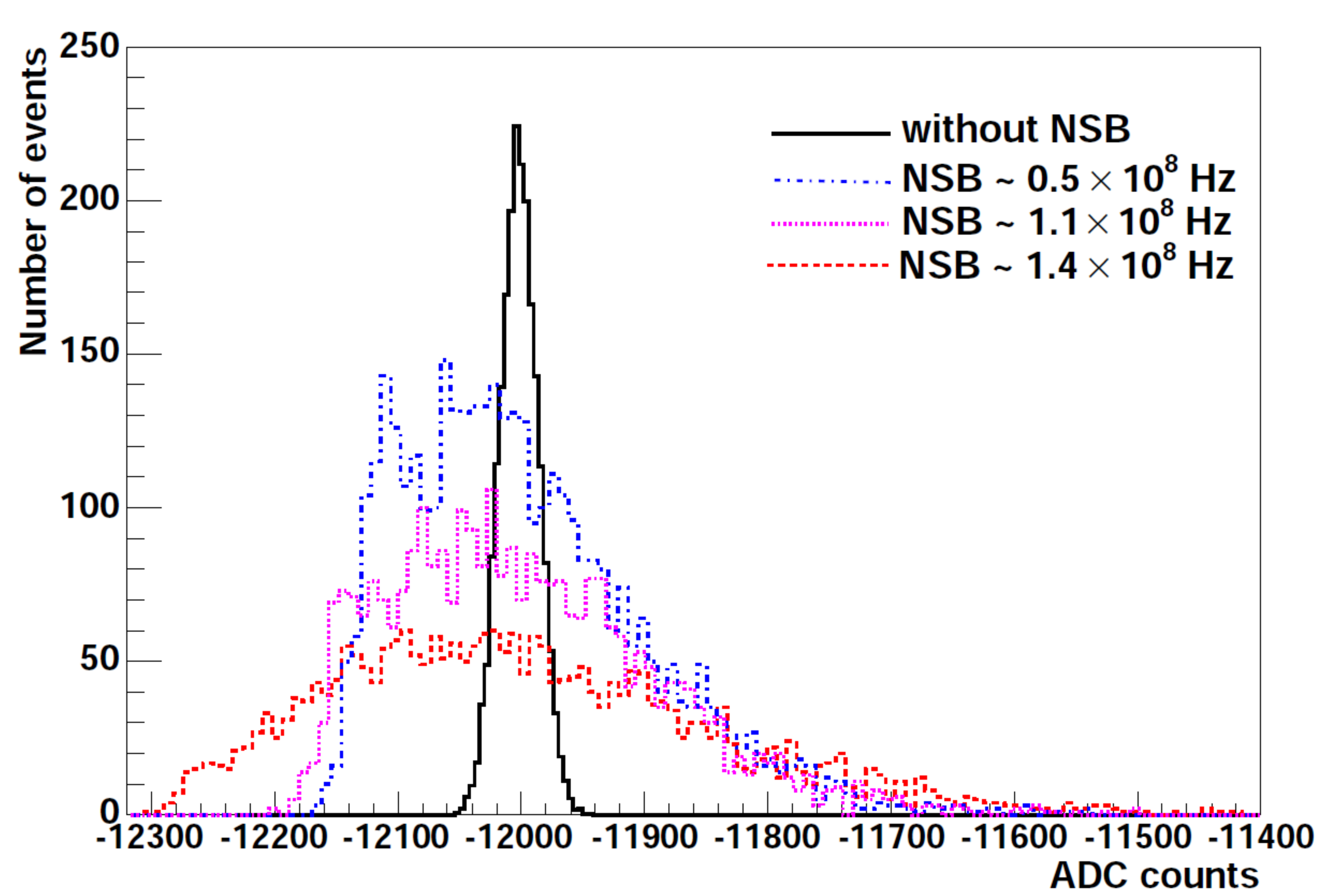}
  \caption{In black, pedestal distribution in ADC counts in the dark. In color, pedestal distributions for different night sky background (NSB) rates (usual rates in Namibia). An increase in NSB rate induces an increase of the pedestal width.}
  \label{fig:pedestal}
 \end{figure}

\subsection{Pedestal and night sky background}
 
The pedestal position is defined as the mean ADC value recorded in absence of Cherenkov light. Pedestals are determined during observation runs as often as possible (approximately every minute). They are estimated on each pixel that doesn't contain Cherenkov light.

The night sky background increases the width of the pedestal distribution but the pedestal mean doesn't vary, as shown in the figure \ref{fig:pedestal} for usual rate in Namibia.

\subsection{Flat-Fielding}

Flat-Field coefficients are used to determine the inhomogeneity between the different pixels. It can be due to different quantum efficiencies of the PMTs and different reflectivities of the Winston cones. These flat-field coefficients are measured during specific flat-field runs, during which cameras are uniformly illuminated with LED flashers mounted on the telescope dish. Flat-field runs are performed as often as possible when the weather does not allow observation runs.

\begin{figure}[ht]
  \centering
  \includegraphics[width=0.42\textwidth]{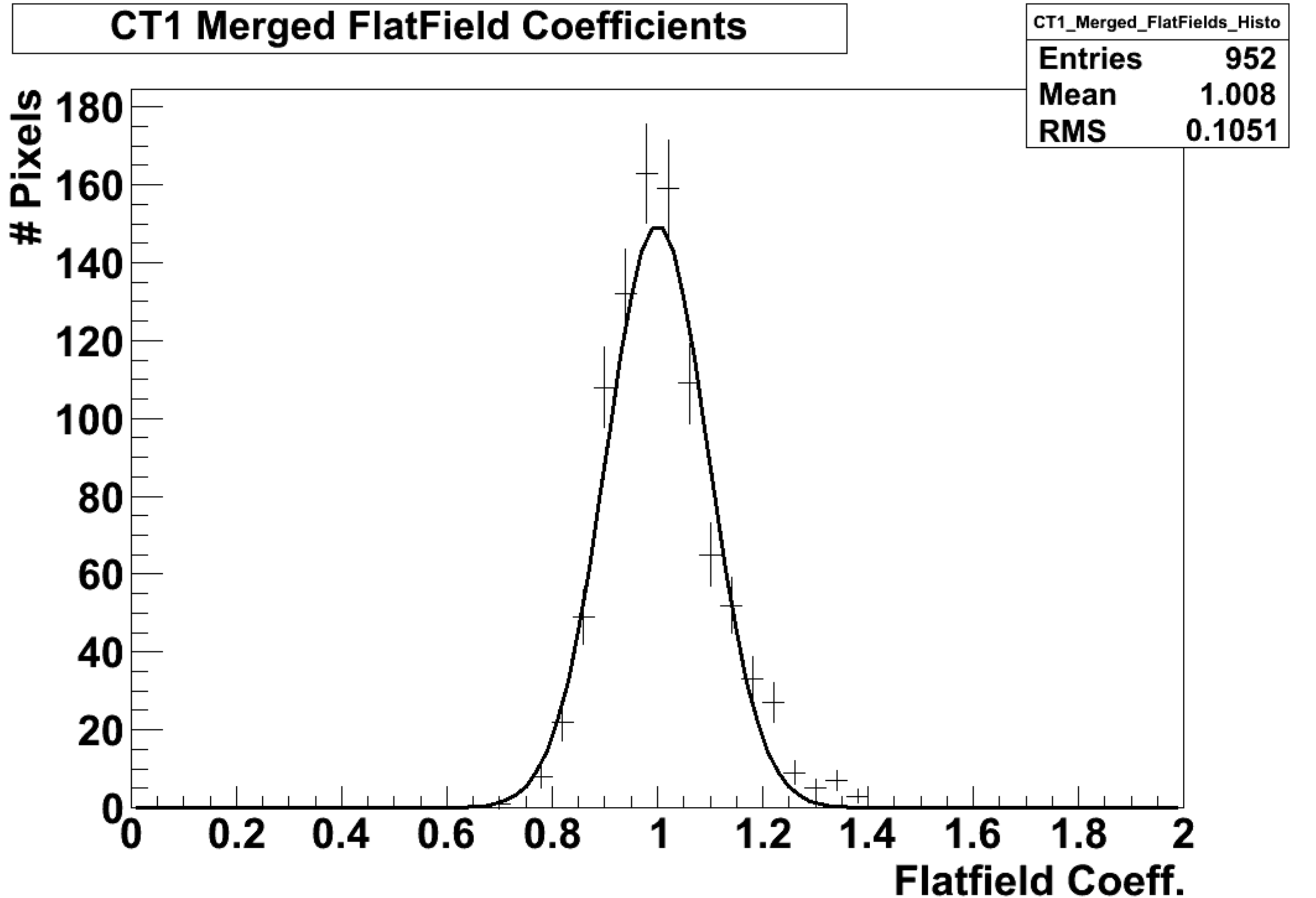}
  \caption{Distribution of the flat-filed coefficients of the pixels of a H.E.S.S. camera.}
  \label{fig:flatfield}
\end{figure}

The distribution of the flat-field coefficients defined by the ratio of the mean intensity of each pixel and the total mean intensity is shown in the figure \ref{fig:flatfield}.

\section{Muon ring calibration}

Once each of the calibration parameters is determined and the conversion from ADC counts to photoelectrons is well defined, the calibration with muon ring can be carried out. 

The study of muons in H.E.S.S. is required to understand losses of Cherenkov light in the detector. As muons only diffuse a few of their energy during the distance they can be observed by the H.E.S.S. mirrors ($\sim$ 400-600 m), the distribution of Cherenkov photons that they emit can be simply calculated. 

Then, by modeling photon travel through the detector, a comparison can be established between the real muon ring picture and the modeled one, directly leading to the optical efficiency of the detector.

The method of reconstruction of muon rings and the determination of the optical efficiency will be described.

\subsection{Optical efficiency}

During their travel inside the detector an important part of the Cherenkov photons is lost. The percentage of photons remaining in the PMTs is called optical efficiency. It can be divided into several contributions: 
\begin{itemize}[itemsep=0pt,leftmargin=4mm]

\item The reflectivity of H.E.S.S mirrors integrated over wavelength : Re $\sim$ 80\%
\item The shadow of camera and masts : Sh $\sim$ 10\%

\item The collection efficiency, which is determined by the loss of photons between mirrors and camera and by the Winston cones reflectivity : Co $\sim$ 70\%
 
\item The quantum efficiency of the PMTs integrated over wavelength : QE $\sim$ 20\%

\end{itemize} 
The total optical efficiency is thus given by the following expression:
\begin{eqnarray*}
\textrm{Opt. eff. = Re $\times$ (1-Sh) $\times$ Co $\times$ QE $\sim$ 10\%}
\end{eqnarray*}

The above values have been determined for a perfect H.E.S.S. telescope. The degradation of the optical efficiency due to dust and ageing of the material can then be obtained by the muon ring reconstruction.

\begin{figure*}[t]
  \centering
  \includegraphics[width=0.66\textwidth]{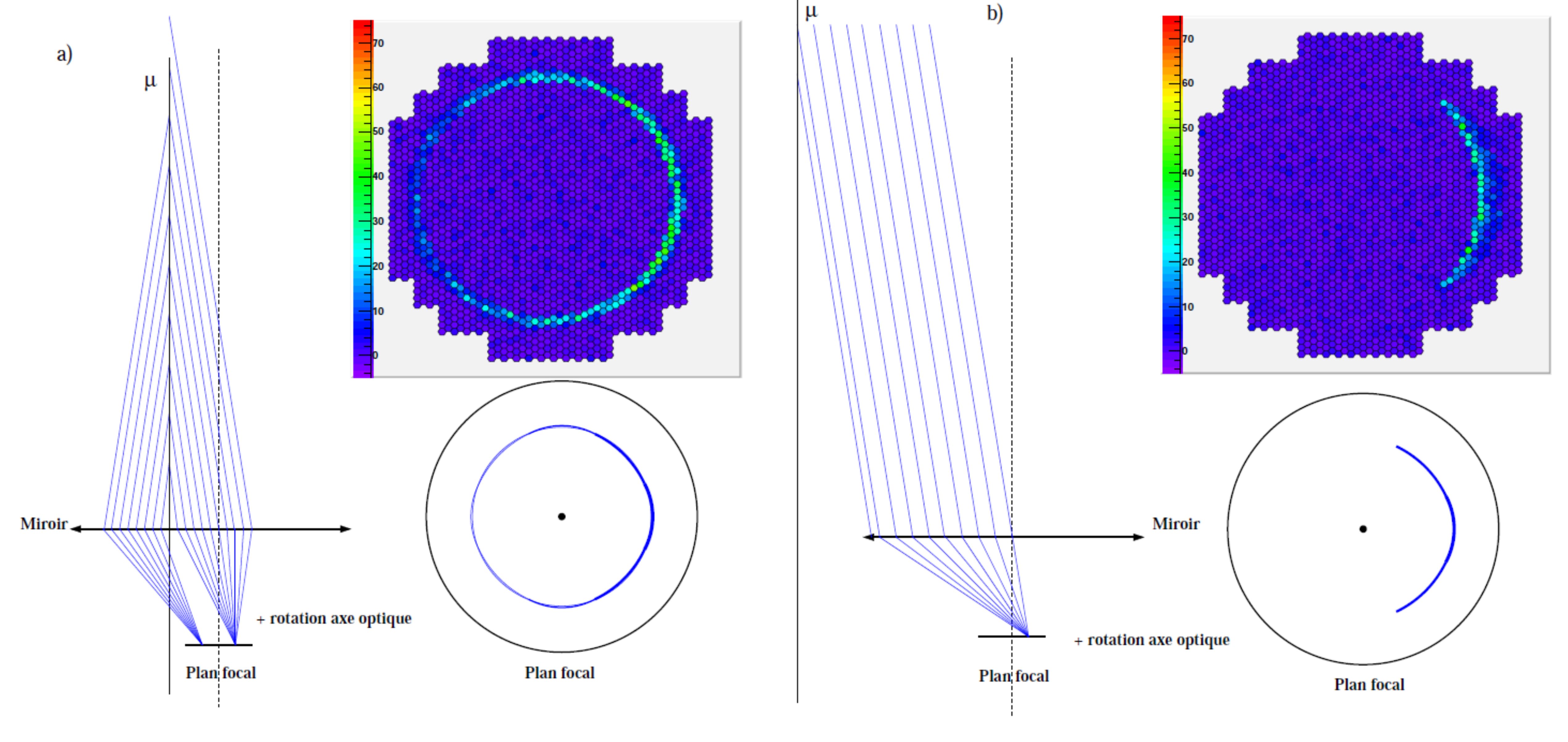}
  \includegraphics[width=0.33\textwidth]{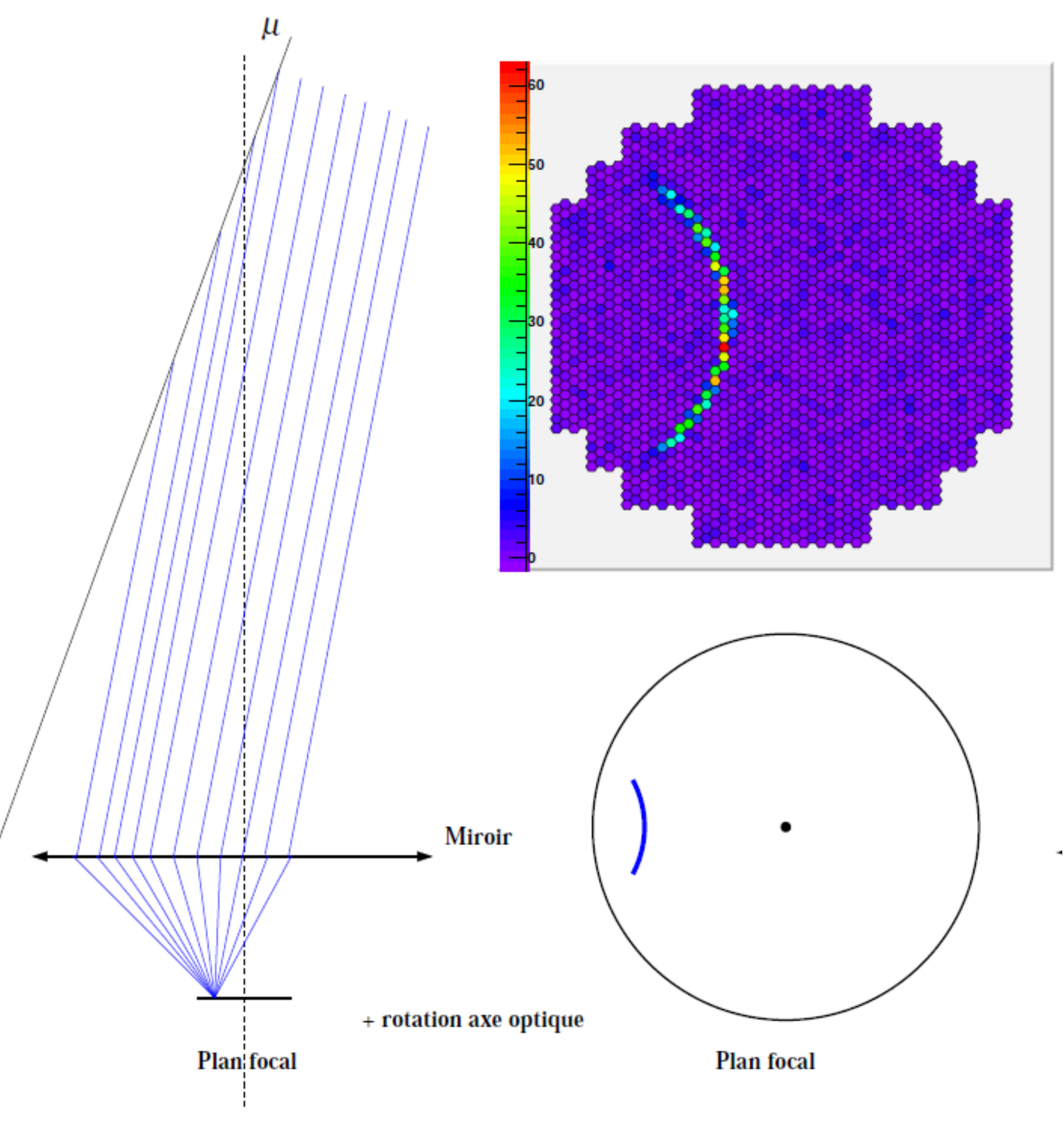}
  \caption{Muon ring behaviour in function of the muon impact parameter and inclination relative to the optical axis. Left: muon falling inside the mirror area. Middle: muon falling outside the mirror area. Right: muon falling outside the mirror area with a non-zero angle with the optical axis.}
  \label{muon_behav}
 \end{figure*}

\subsection{Muon ring model}

The muon ring model for a telescope with a circular mirror is described by Vacanti et al. \cite{bib:vacanti}. It gives the total number of photoelectrons obtained in the whole camera as a function of muon parameters: 
\begin{eqnarray}
  	\frac{d^3N}{dld\phi d\lambda} = \frac{\alpha}{2}\sin{(2\theta_c)}
  	\frac{\psi(\lambda)}{\lambda^2}
  	D(\phi)a(l,\lambda) 
\end{eqnarray}
where $\alpha$ is the fine structure constant, $\theta_C$ is the Cherenkov angle proportional to the ring radius, $\psi(\lambda)$ is the optical efficiency, $l$ is the muon path length and $a(l,\lambda)$ is the attenuation coefficient of the atmosphere. 

$D(\phi)$ is the chord defined by the intersection with the mirror plane and the plane of the photons trajectory, such as: 
\begin{eqnarray}
D(\phi) = \left\{
  \begin{array}{l l}
    2R \sqrt{1-(\rho/R)^2 \sin^2{\phi}} & \quad \textrm{if   $\rho$ $>$ R}\\\\
    R \left(\sqrt{1-(\rho/R)^2 \sin{\phi}^2}\right. \\
     \qquad\qquad  +\left.(\rho/R)\cos{\phi}  \vphantom{\sqrt{}} \right) &  \quad \textrm{if   $\rho$ $\leq$ R}
  \end{array} \right.
\end{eqnarray}
where $\rho$ is the muon impact parameter, $R$ is the mirror radius and $\phi$ is the azimuth angle in the focal plane.

The figure \ref{muon_behav} shows the muon ring behaviour as a function of the muon impact parameter and its inclination with the optical axis. The ring is complete when the muon falls inside the mirror plane; otherwise it's an arc. 
 The inclination with the optical axis displaces the center of the ring.

The expression which is used by the H.E.S.S. collaboration for muon ring reconstruction gives the number of photoelectrons inside a given PMT \cite{bib:rovero}:
\begin{eqnarray}\label{model_PMT}
  	\frac{d^4N}{drd\phi d\lambda d\theta} & = & \frac{\alpha}{2}\sin{(2\theta_c)}
  	\frac{\psi(\lambda)}{\lambda^2}
  	D(\phi)a(r,\lambda) \nonumber \\ 
& & \times \frac{\exp\left(-\frac{(\theta-\theta_c)^2}{2\sigma_T^2(r,\theta_c)}\right)}{\sqrt{2\pi}\sigma_T(r,\theta_c)} 
\end{eqnarray}
where $r$ is the distance from the center of the dish to a given mirror, $\theta$ is the angular distance from the center of the muon ring to a given point in the studied PMT and $\sigma_T$ is the width of the ring assumed to be gaussian in this study.

 \begin{figure}[ht]
  \centering
  \includegraphics[width=0.42\textwidth]{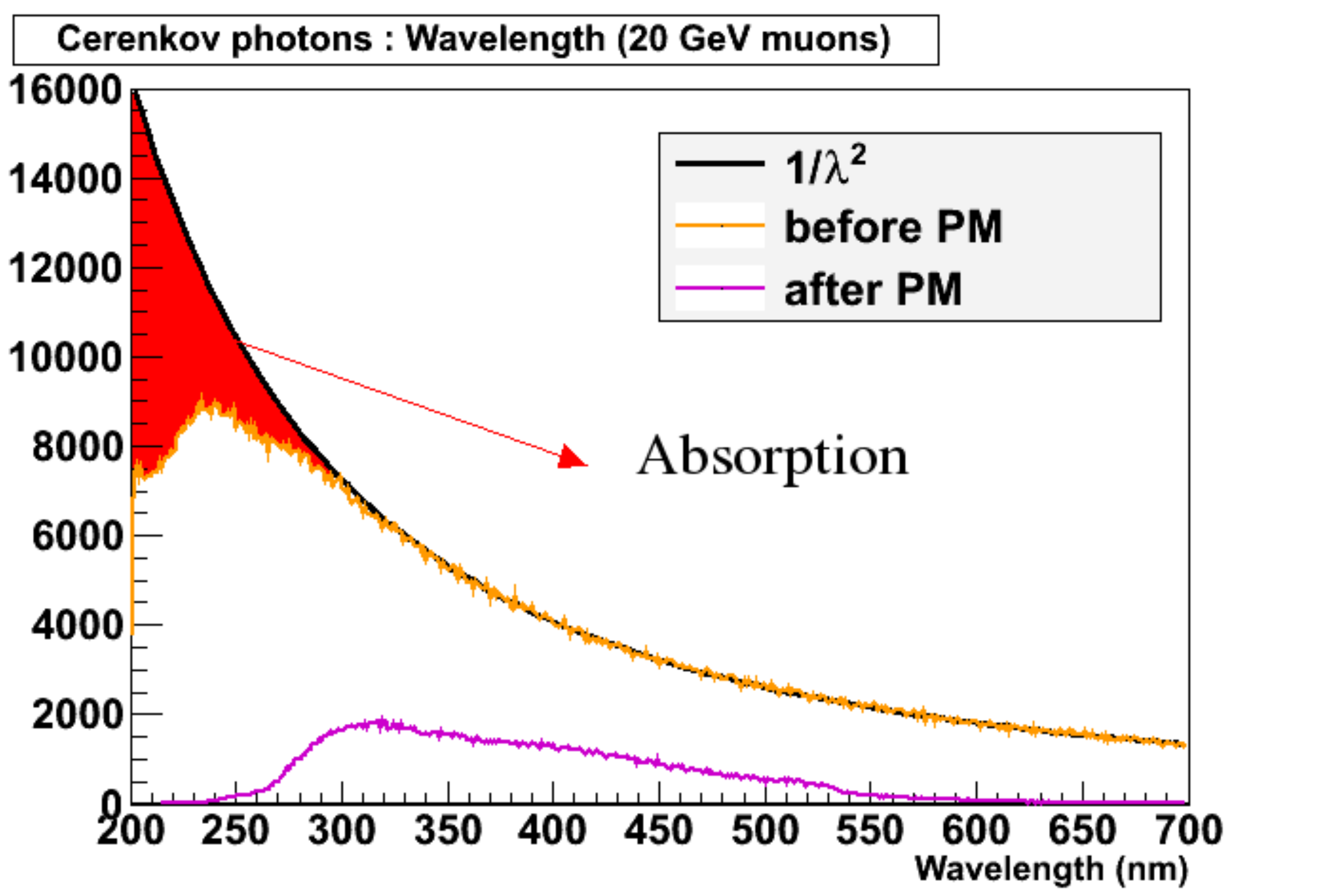}
  \caption{Distribution of Cherenkov photons at the ground (in yellow) normalized to fit with a $1/\lambda^2$ function (in black). The purple curve represents the distribution of photoelectrons after passing through the PMTs }
  \label{fig:absorption}
 \end{figure} 

The attenuation of the atmosphere is reproduced by Monte Carlo simulations. Muons are launch in the atmosphere and the spectrum of Cherenkov photons received at the ground, presented in yellow in the figure \ref{fig:absorption}, is integrated over wavelength. The integration range is taken between 270 and 700 nm where the PMT quantum efficiency is significant and where the attenuation is limited.

\subsection{Muon efficiency reconstruction}

Two main steps are needed to reconstruct the muon ring and determine the telescope optical efficiency. 

A computation using the Karim\"aki algorithm \cite{bib:karimaki} on a cleaned image is firstly performed to determine the radius and the position of the ring center and used to do a first muon selection.

In a second step, the equation (\ref{model_PMT}) is fitted to the intensity of each pixel of the cleaned image via a maximum-likelihood method. From this fit, the different parameters of the muon and the muon efficiency are extracted.

\begin{figure}[h]
  \centering
  \includegraphics[width=0.42\textwidth]{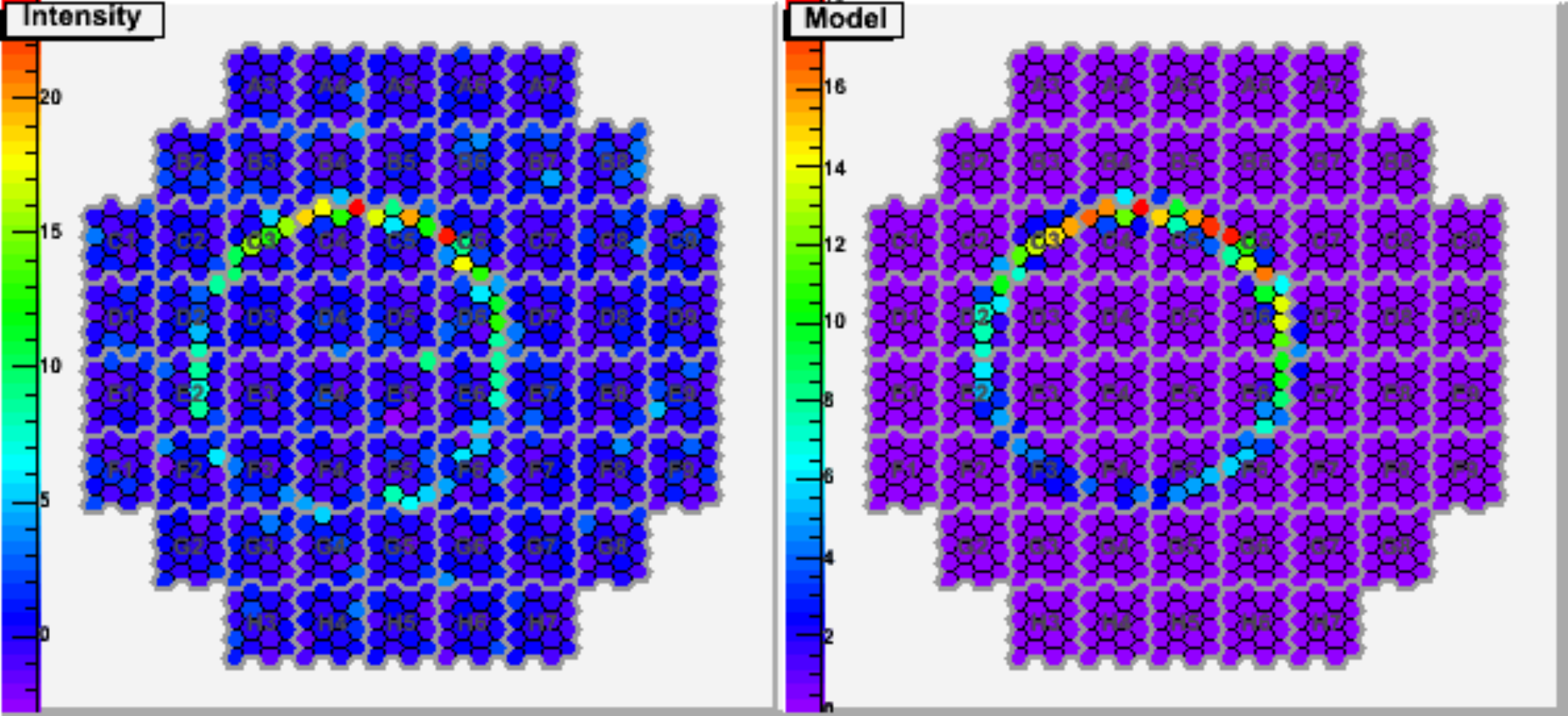}
  \caption{Left: real muon ring image in a H.E.S.S telescope. The intensity are expressed in photoelectron counts. Right: fitted muon ring model obtained with the equation (\ref{model_PMT}). Colour scales are different in the two pictures.}
  \label{muon_model}
 \end{figure}

An example of a muon observed during a H.E.S.S. run and its fitted model is shown in figure \ref{muon_model}.


In this method, all pixels with an intensity greater than a threshold participate to the adjustment, allowing to reject an image ring contaminated by hadron shower, even faintly.

 \begin{figure*}[t]
  \centering
  \includegraphics[width=0.49\textwidth]{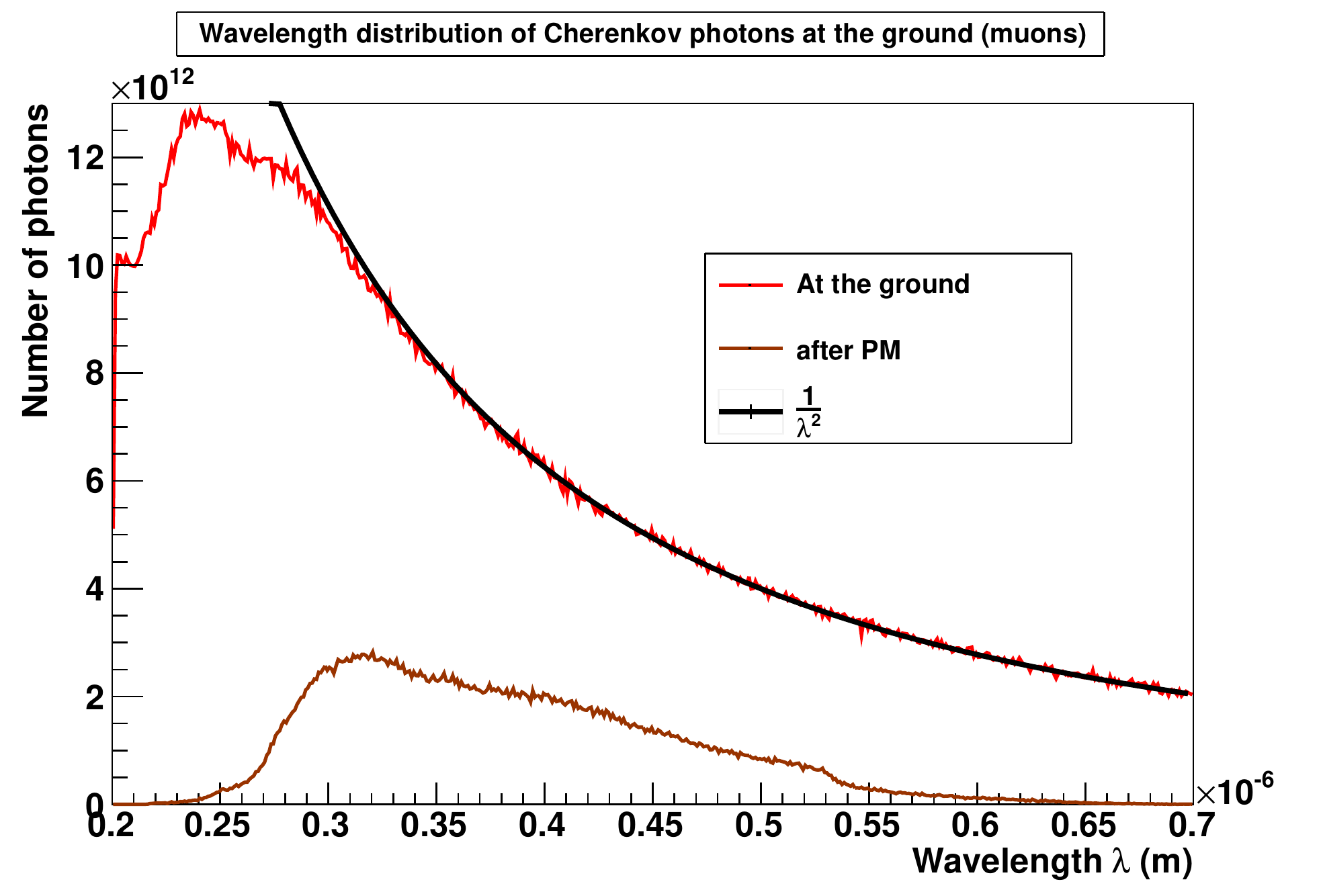}
  \includegraphics[width=0.49\textwidth]{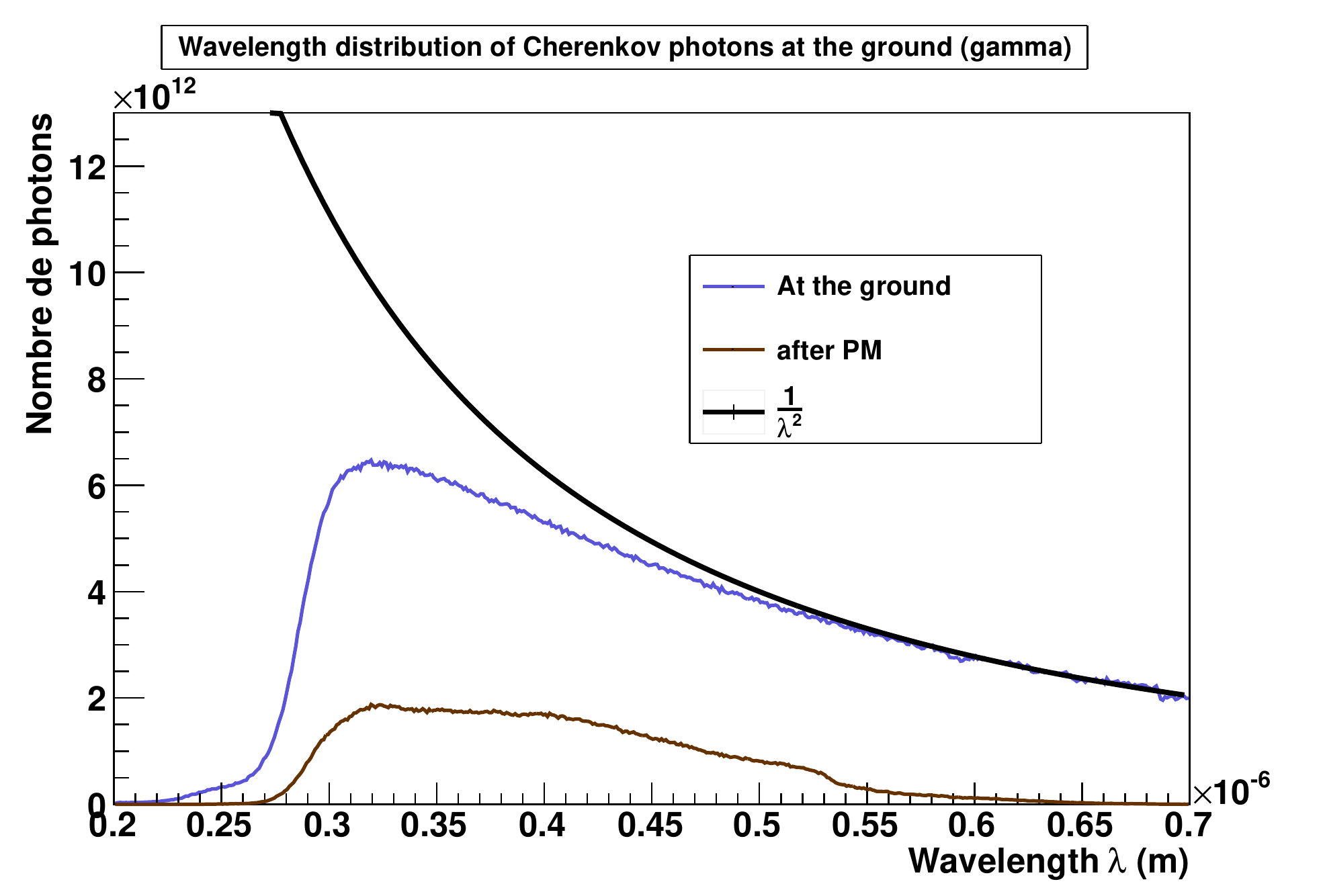}
  \caption{Simulated distributions of Cherenkov photons at the ground normalised to fit with a $1/\lambda^2$ function (in black). Left: Cherenkov photons emitted by muons. Right: Cherenkov photons emitted by 500 GeV gamma-rays. The brown curves represent the same distributions of photons after passing through the photomultiplier tubes.}
  \label{muon_gamma}
 \end{figure*}

  \begin{figure*}[b]
  \centering
  \includegraphics[width=\textwidth]{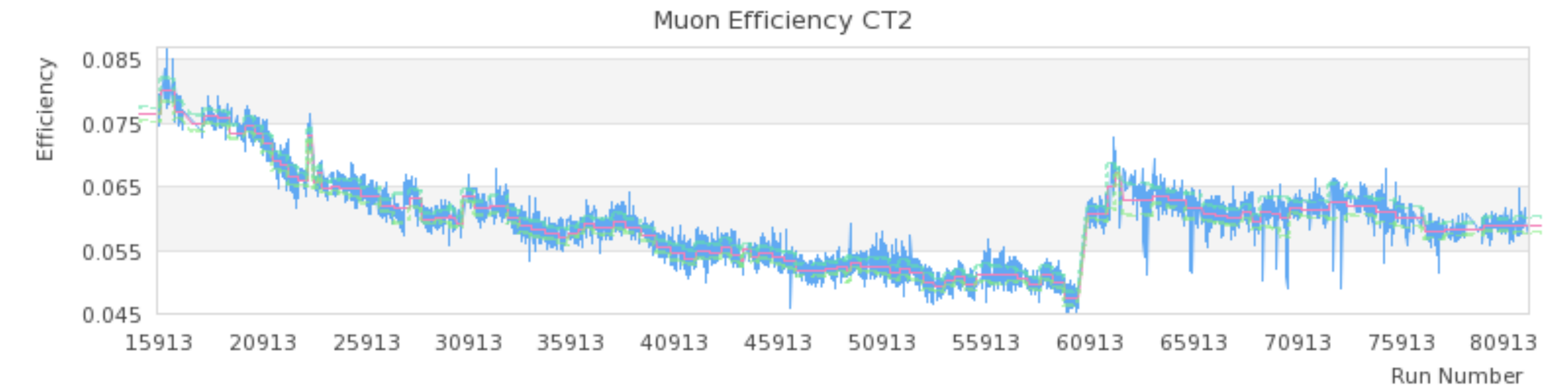}
  \caption{Optical efficiency of a H.E.S.S. telescope as a function of the run number.}
  \label{CT2_efficiency}
 \end{figure*}

\subsection{From muon to gamma}

The optical efficiency is wavelength dependent. This effect induces a difference when looking at gamma-rays compared with muons. 

The figure \ref{muon_gamma} shows the distributions of Cherenkov photons at the ground emitted by gamma-rays on the one hand and by muons on the other hand. For gamma-rays, a large part of the spectrum is attenuated during the travel through the atmosphere.

Due to the differences of these spectra, the optical efficiency of the telescope will differ between these two kinds of particles.

In order to recover the optical efficiency for gamma-rays from the muon ring reconstruction, a conversion factor must be applied to the muon efficiency. This factor is obtained by dividing the integral over wavelength of each of the distributions shown in the figure \ref{muon_gamma}. It is then directly applied to the muon efficiency achieved by muon ring reconstruction.

\subsection{Reference efficiency}

To understand the degradation of a H.E.S.S. telescope and take it into account in the reconstruction of gamma-ray energy and flux, a reference efficiency has to be established from a Monte Carlo study. 

Muons are launched on a simulated perfect telescope and the reconstruction procedure of muon rings is applied to them. The reference efficiency is the efficiency obtained by this procedure.

During an observation run, the reconstructed efficiency is then divided by the reference efficiency and the resulting factor is applied to the gamma-ray energy reconstruction.

The optical efficiency of a H.E.S.S. telescope over time is shown in the figure \ref{CT2_efficiency}. This graph shows the degradation of the telescope. One can also see several small jumps in the efficiency which are correlated with the increase of the high voltage supply in order to recover the nominal gain value. The big jump around the run 61000 is due to the replacement of the telescope mirrors.

\section{Conclusions}

The calibration parameters of the H.E.S.S. cameras are well reconstructed and monitored over years. Two independent calibration patterns are in operation in the H.E.S.S collaboration and have shown that the intensity in photoelectrons can be evaluated with a precision better than 5 \%.

Calibration with muon ring allows to fully understand the losses of Cherenkov photons in the detector. It needs a Monte Carlo study which induces a good knowledge of the detector itself and of the atmosphere. Especially, a precise understanding of the attenuation of Cherenkov light in the atmosphere, of the development of electromagnetic showers and propagation of muons in the atmosphere is required.

\vspace*{0.5cm}

\end{document}